# Classifying Brazilian Monetary Policy States with Interest Rate Gaps and Markov Models

Gabriel de Macedo Santos gabriel.macedo@sou.inteli.edu.br

Instituto de Tecnologia e Liderança

**Abstract**

This paper develops a monthly framework for classifying Brazilian monetary policy that distinguishes the level of monetary restraint, changes in that stance, and statistical regimes associated with Selic adjustments. The framework combines a seven-state deterministic classifier, an empirical Markov chain, a Gaussian hidden Markov model, and Markov-switching regressions. Project outputs provide 145 classified observations from May 2014 to May 2026, with a partial final month. Under the implemented nominal neutral rate convention, restrictive states account for 88.3% of observations. A four-state HMM achieves a lower BIC, whereas the three-state model records a lower temporal validation loss. Switching regressions select three Selic adjustment regimes after solutions with degenerate variances are excluded. An additional sensitivity exercise reduces the restrictive share to 51.0% when the neutral benchmark is interpreted as real, demonstrating how strongly the results depend on benchmark measurement. The contribution is an auditable comparative application for monetary monitoring. The findings describe historical patterns conditional on the data and specifications; they do not establish predictive superiority or causal identification.

## 1 Introduction

Assessing monetary policy requires distinguishing the position of interest rates relative to an equilibrium benchmark from the direction in which rates are moving. A Selic rate cut can occur while the policy rate remains above its neutral benchmark. Similarly, an unchanged nominal rate can coincide with a change in monetary stance if the estimated neutral rate shifts. Classifications based solely on rate increases, cuts, or holds do not preserve these distinctions.

This paper organizes these dimensions into a monthly framework applied to Brazil. The central question is how to represent the state of monetary policy and its transition dynamics without conflating monetary restraint with changes in the policy instrument. The starting point is the gap between the Selic rate and a neutral rate benchmark, combined with the three-month change in that gap. The resulting rule produces observable, economically interpretable states. A Markov chain summarizes historical transitions between them. In parallel, a hidden Markov model, or HMM, identifies latent segments in a set of macrofinancial variables, while Markov-switching regressions describe patterns in Selic adjustments.

The contribution is applied. The paper does not introduce a new regime estimator or a new structural measure of neutrality. Instead, the project brings established methods together in a procedure that makes the path from inputs to classification traceable and allows comparison of what each model actually identifies. This organization makes three distinct questions explicit: how restrictive interest rates are relative to the benchmark, which states tend to follow one another in the sample, and which patterns emerge in the distribution of observables and in Selic adjustments.

The outputs preserved in the project show a predominance of restrictive states under the implemented nominal neutral rate convention. This conclusion, however, depends on a source characteristic that remains undocumented: the neutral rate file provides a series named *r_star_mean_all* without its generating equation or an unambiguous identification as a nominal or real rate. A sensitivity exercise addressing this interpretation substantially changes the composition of states. Measuring neutrality is therefore central to the analysis rather than an assumption that can remain implicit.

A further distinction is needed between historical classification and forecasting. Smoothed probabilities from the latent models use the full sample, and the activity gap employed in the project comes from a two-sided filter. Moreover, the files do not contain a recurring real-time evaluation of transition forecasts or Copom decisions. The paper therefore examines the descriptive usefulness and limitations of the framework without treating retrospective fit as demonstrated predictive performance.

## 2 Related literature

The monetary policy rule literature provides a reference for distinguishing the level of interest rates from their response to macroeconomic conditions. Taylor (1993) formalizes a rule relating the interest rate to inflation and activity. For Brazil, Minella et al. (2003) examine the central bank's reaction function and emphasize the role of inflation expectations. These contributions motivate the inclusion of inflation, activity, and exchange rates in the regression block without implying that the implementation reproduces their identification strategies.

Comparison with neutrality involves a variable that is not directly observed. Laubach and Williams (2003) estimate the natural real interest rate through an unobserved components model and emphasize measurement uncertainty. In the present application, the neutral rate is an input supplied to the project rather than a variable reestimated within it; the series is taken from Santos (2026), who combines statistical trends, market prices, and state-space models in an ensemble that tracks Brazil's real neutral rate. The distinction between real and nominal rates must therefore be resolved before strong economic interpretations can be attached to gap levels.

Hamilton (1989) provides a foundational framework for modeling discrete regime changes in macroeconomic time series. In such models, the state is latent and evolves according to a Markov chain. The Gaussian HMM used here follows the same principle of temporal dependence between states, with a multivariate emission distribution. Expectation-maximization estimation and inference follow the general framework described by Rabiner (1989). In contrast, the classifier's empirical chain starts from states that are observed by construction; transition probabilities are estimated from counts, without inferring hidden states.

Sims and Zha (2006) demonstrate the importance of distinguishing changes in policy coefficients from changes in shock variances. Davig and Leeper (2007) show that interpreting the Taylor principle under regime switching also depends on the dynamics across regimes. These references motivate two cautions: regimes with different intercepts do not establish changes in the response to inflation, and comparing individual coefficients does not, by itself, establish macroeconomic equilibrium determinacy.

Two further issues connect the application to the methodological literature. The first concerns the sensitivity of the cyclical component produced by the Hodrick-Prescott filter. Ravn and Uhlig (2002) provide a basis for adjusting the smoothing parameter to the observation frequency, while Hamilton (2018) discusses the filter's dynamic distortions and endpoint problems. The second concerns model selection: Schwarz's (1978) BIC penalizes complexity but does not replace checks for degenerate solutions, verification of input data, or appropriate temporal evaluation.

## 3 Data and sample construction

### 3.1 Sources and transformations

The supplied dataset combines a spreadsheet of financial and macroeconomic series, a spreadsheet of inflation expectations at the relevant policy horizon, and a file containing the average of neutral rate estimates produced by the multi-block ensemble framework of Santos (2026). The first spreadsheet provides Bloomberg identifiers for the Selic rate, 252- and 294-business-day DI swaps, the USDBRL exchange rate, 12-month-ahead expected IPCA inflation, observed 12-month IPCA inflation, and the level of the IBC-Br activity index. This paper uses the local files supplied with the project, without replacing their observations with later data extractions.

Monthly standardization takes the last available nonmissing value of each series within a month and assigns the record to the month's final calendar day. This does not mean that all variables were observed on the same release date. The inflation series at the relevant policy horizon has an irregular frequency and is carried forward until a new observation becomes available. Its monthly changes can therefore reflect updates to a projection that remained unchanged between releases.

**Table 1:** Coverage of the estimation samples

| Block | Start | Final monthly label | Observations |
|---|---|---|---|
| Consolidated monthly dataset | Apr. 2004 | May 2026 | 266 |
| Available neutral benchmark | Feb. 2014 | May 2026 | 148 |
| Rule and Markov chain | May 2014 | May 2026 | 145 |
| Gaussian HMM | May 2014 | May 2026 | 145 |
| Switching regressions | Mar. 2008 | May 2026 | 219 |

*Source: project input files and outputs. May 2026 is a partial month. Sample differences reflect each block's required variables; constructing the direction of the gap removes three additional months from the rule's sample.*

The neutral rate file begins in February 2014. The three-month change requirement makes May 2014 the first classified month. The HMM uses the same 145 months because its variables include the current gap and its change. The regressions do not require the neutral benchmark: availability of inflation at the relevant horizon permits 219 observations starting in March 2008. Consequently, fit statistics across these three blocks do not represent a direct comparison on a common sample.

### 3.2 Information dates and measurement

The last financial spreadsheet observation is dated May 6, 2026. The standardized record labeled May 31 therefore represents a month still in progress. The final observation of inflation at the relevant horizon is dated May 3. The neutral series has a final reference month of May but does not provide a publication date. The financial data cutoff is known; a common, verifiable real-time information set across all series is not documented.

The IBC-Br is supplied as an index level even though the spreadsheet header refers to year-over-year growth. Its values and use in the code correspond to a level index transformed into logarithms. The file does not fully document the seasonal treatment of the series. The HP-filtered component is therefore referred to as an activity gap proxy, without equating it to a structural estimate of the output gap. Repeated recent index values may reflect the retention of the last published observation, which matters for endpoint interpretation.

Although the neutral series' generating methodology is documented in Santos (2026), the supplied material lacks its' uncertainty interval, historical data vintages, and a complete record of software dependency versions. These omissions do not prevent verification of the classifications and statistics in the files, but they limit independent reconstruction of the information that would have been available at each historical date.

## 4 Methodological framework

### 4.1 Interest rate gaps and the activity proxy

Let $i_t$ denote the Selic rate and $n_t$ the neutral benchmark treated by the project as nominal, both expressed as annual percentages. The current stance and its direction are defined as

$$g_t = i_t - n_t, \qquad d_t = g_t - g_{t-3}. \qquad (1)$$

Gaps are measured in percentage points. By definition, $d_t = (i_t - i_{t-3}) - (n_t - n_{t-3})$. A falling gap therefore does not require a Selic cut: an increase in the neutral benchmark also reduces it. The terms "rising" and "falling" in the classification refer to the gap rather than exclusively to the policy instrument.

For the 252-business-day DI swap rate, denoted by $j_t$, the forward gap and the slope relative to the Selic rate are defined as

$$f_t = j_t - n_t, \qquad s_t = j_t - i_t, \qquad f_t = g_t + s_t. \qquad (2)$$

The term forward gap is an operational convention. Swap rates incorporate expectations and premia; their difference from the neutral benchmark does not separately identify the expected monetary policy path. The final identity in equation (2) is relevant to the assessment of the HMM's variable set.

With 12-month expected inflation $\pi_t^e$, the ex ante real rate and the neutral benchmark converted into real terms follow the Fisher relation:

$$r_t = 100\left(\frac{1+\frac{j_t}{100}}{1+\frac{\pi_t^e}{100}} - 1\right), \qquad r_t^* = 100\left(\frac{1+\frac{n_t}{100}}{1+\frac{\pi_t^e}{100}} - 1\right). \qquad (3)$$

It follows that $r_t - r_t^* = \frac{f_t}{1+\frac{\pi_t^e}{100}}$. The real gap contains nearly the same information as the nominal forward gap and is excluded from the baseline HMM specification. Because the deflator varies over time, their correlation need not be exactly one.

The activity gap proxy uses the logarithm of the IBC-Br level, $z_t = \ln Y_t$, and the HP trend $\tau_t$:

$$\min_{\{\tau_t\}} \left\{ \sum_{t=1}^{T} (z_t - \tau_t)^2 + \lambda \sum_{t=2}^{T-1} (\tau_{t+1} - 2\tau_t + \tau_{t-1})^2 \right\}, \qquad y_t = 100(z_t - \tau_t). \qquad (4)$$

The smoothing parameter is $\lambda = 129\,600$, following the frequency adjustment in Ravn and Uhlig (2002). The factor of 100 is applied to the cycle after filtering. This order preserves the relative scale of the minimization problem and matches the implementation examined here.

### 4.2 Deterministic classification

The rule uses a level band of $\theta = 0.50$ percentage points and a direction band of $\phi = 0.25$ percentage points. Observations with $g_t > \theta$ are restrictive, those with $g_t < -\theta$ are accommodative, and those with $|g_t| \leq \theta$ are neutral. Outside the neutral band, direction is rising if $d_t > \phi$, stable if $|d_t| \leq \phi$, and falling if $d_t < -\phi$. These combinations produce seven states because the neutral region is aggregated into a single transition category.

Months without a gap or a direction remain unclassified. The neutral category is not used to fill missing data. Thresholds are operational choices in the project, not estimated parameters or bands derived from standard errors. The high, medium, or low confidence indicator exported by the code measures distance from the rule's boundaries; it is not a statistical probability of correct classification.

### 4.3 Markov chain for observed states

Let $R_t$ denote the state defined by the rule. A homogeneous first-order chain is assumed as a descriptive approximation. With $N_{ab}$ transitions from state $a$ to state $b$, the estimator is

$$\hat{p}_{ab} = \frac{N_{ab}}{\sum_c N_{ac}}, \qquad \hat{P} = (\hat{p}_{ab}). \qquad (5)$$

The procedure considers only consecutive monthly pairs. The classified sample contains 144 transitions across 145 observations. Rows represent origin states and columns represent destination states. For a state with a persistence probability below one, the implied geometric duration, as in the Markov chain framework discussed by Hamilton (1994), is

$$\widehat{D}_a = \frac{1}{1 - \hat{p}_{aa}}. \qquad (6)$$

This duration is a property of the estimated model, distinct from the observed average episode length, particularly at the sample endpoints. The stationary distribution $\omega$ satisfies $\omega^\top \hat{P} = \omega^\top$ and $\sum_a \omega_a = 1$. Interpreting it as long-run occupancy requires an unchanged transition matrix and the relevant ergodicity conditions. It is not interpreted as an unconditional forecast for Brazilian monetary policy over arbitrary periods.

The project exports a chi-squared test of independence between origin and destination states and adjusted cell residuals, following the usual contingency table construction in Haberman (1973). Table sparsity calls for caution. Moreover, serial dependence does not, by itself, verify the sufficiency of a first-order assumption or temporal homogeneity. The three-month change used by the rule also contains overlapping information across months, mechanically contributing to state persistence.

### 4.4 Gaussian hidden Markov model

The HMM uses a standardized vector $x_t$ containing seven variables: the current gap, its three-month change, the forward gap, the three-month change in inflation at the relevant horizon, the yield curve slope, the three-month log change in USDBRL, and the activity gap proxy. For a latent state $H_t \in \{1, \ldots, K\}$,

$$\Pr(H_t = b \mid H_{t-1} = a) = A_{ab}, \qquad x_t \mid H_t = k \sim \mathcal{N}(\mu_k, \Sigma_k). \qquad (7)$$

The covariance matrix $\Sigma_k$ is diagonal. With $d = 7$ variables, the nominal number of free parameters is

$$q_K = (K-1) + K(K-1) + 2Kd, \qquad \mathrm{BIC}_K = q_K \ln T - 2\ell_K. \qquad (8)$$

Models with $K = 3$ and $K = 4$ are compared using seven initializations, up to 500 iterations, and a tolerance of $10^{-4}$. The procedure does not simply take the lowest BIC across all initializations: within each $K$, a score combines BIC, a penalty for small states, and validation loss; across specifications, it adds BIC dispersion across seeds and complexity per observation. In aggregate, the exported score is

$$J_K = \mathrm{BIC}_K + 0.30\,\mathrm{sd}(\mathrm{BIC}_{K,b}) + 35\frac{q_K}{T} + 2T_v \mathrm{med}_b(\mathrm{NLL}_{K,b}). \qquad (9)$$

The index $b$ denotes random seeds. Models with a minimum state share below 8% are penalized in selection. The score's weights are heuristic: they do not represent a posterior distribution or an established information criterion. Their use must be distinguished from conventional BIC.

Validation reserves the final 29 months, from January 2024 to May 2026, and reestimates the models on the preceding 116 months. Standardization is fitted on training data alone. Loss is obtained by evaluating the final block as a sequence initialized with the reestimated model's initial distribution, without explicitly propagating the filtered distribution from the last training month. Furthermore, the activity gap previously computed over the full sample remains in the variable set. This is therefore a partial temporal split rather than a fully real-time evaluation.

Exported probabilities are smoothed, $\Pr(H_t = k \mid x_{1:T})$, whereas the state sequence is obtained through Viterbi decoding. These outputs solve different inference problems and may occasionally disagree. Contemporaneous monitoring would require filtered probabilities and estimation that respects the historical availability of the data.

### 4.5 Markov-switching regressions

The third block considers Gaussian regressions with latent Markov states of the form

$$v_t = c_{M_t} + \beta_{M_t}^{\top} w_t + \varepsilon_t, \qquad \varepsilon_t \mid M_t \sim \mathcal{N}(0, \sigma_{M_t}^2). \qquad (10)$$

The dependent variable $v_t$ is either the Selic rate or its monthly change. Regressors include combinations of the deviation of expected inflation from its target, the activity gap proxy, the exchange rate change, lagged Selic, and the yield curve slope. Appendix A summarizes the six specifications, each estimated with two and three regimes.

In level equations with inertia, a long-run response to inflation can be written as $\frac{\beta_{\pi,k}}{1-\rho_k}$ when the parameterization and stability conditions support that interpretation. Only specification F allows all coefficients to vary across regimes. Model C, selected in the project outputs, switches only the intercept of the first-difference equation. Its states describe conditional patterns of Selic adjustment; they do not establish changes in the central bank's structural response to inflation.

The inflation gap uses the projection at the relevant policy horizon and a target interpolated between the current and following years:

$$\pi_t^{*,12m} = \frac{12 - m_t}{12} \pi_{a(t)}^* + \frac{m_t}{12} \pi_{a(t)+1}^*. \qquad (11)$$

Annual target centers follow the schedule incorporated into the project, referenced to the Central Bank of Brazil's historical record. The interpolation is an approximate convention for a rolling 12-month horizon. The relevant-horizon inflation input, however, does not identify each projection's horizon. Without this metadata, alignment between the projection and interpolated target cannot be established. This limitation affects the inflation regressor's interpretation, although its numerical construction remains reproducible.

Solutions are checked for convergence, regime occupancy, and variance degeneracy. The project discards models whose minimum variance is below $10^{-3}$ squared percentage points or whose variance ratio exceeds $10^{6}$. This check seeks to prevent a regime from fitting months of unchanged Selic with a near-zero variance and generating an artificially high likelihood. It is a numerical admissibility filter, not an estimated regularization procedure or proof of correct specification.

## 5 Empirical results

### 5.1 Composition and persistence of observed states

Under the implemented nominal convention, restrictive states account for 128 of 145 observations, or 88.3% of the sample. There are 35 state changes. The restrictive and rising category contains 56 months, and restrictive and falling contains 53. Distinguishing direction captures an important dimension of variation that would be lost by classifying only the gap's sign.

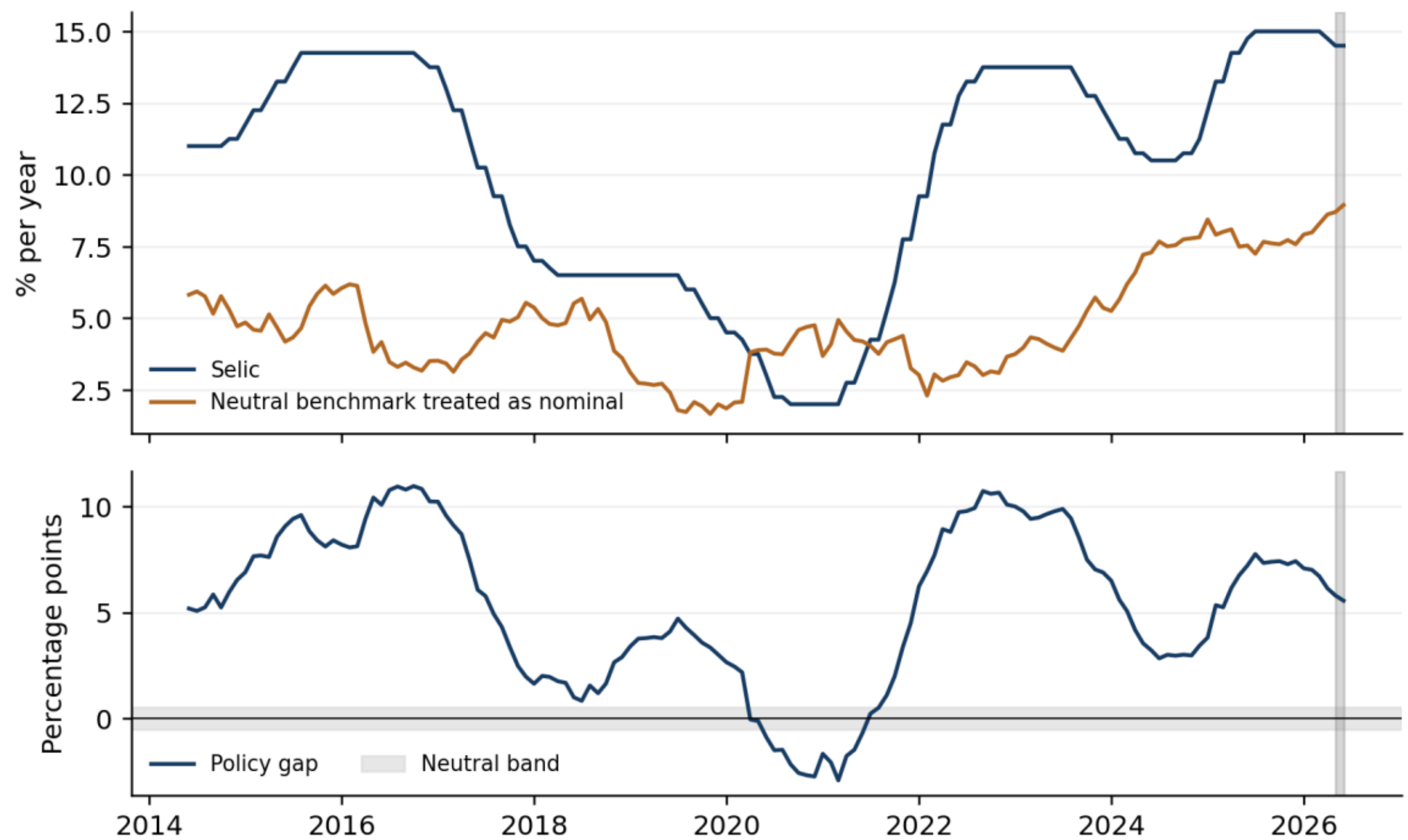

***Figure 1:*** *Selic rate, neutral benchmark, and policy gap*

*Source: calculations from the processed dataset. The neutral benchmark is treated as nominal, following the code. The gray band in the lower panel represents ±0.50 percentage points; the final month is partial. The plotted path does not resolve the economic nature of the original neutral series.*

**Table 2:** State frequencies, persistence, and implied duration

| State | Months | Share | Persistence | Duration in months |
|---|---|---|---|---|
| Neutral transition | 4 | 2.8% | 0.500 | 2.00 |
| Restrictive and rising | 56 | 38.6% | 0.857 | 7.00 |
| Restrictive and stable | 19 | 13.1% | 0.316 | 1.46 |
| Restrictive and falling | 53 | 36.6% | 0.846 | 6.50 |
| Accommodative and rising | 4 | 2.8% | 0.500 | 2.00 |
| Accommodative and stable | 2 | 1.4% | 0.500 | 2.00 |
| Accommodative and falling | 7 | 4.8% | 0.857 | 7.00 |

*Source: Markov block outputs. Persistence is the estimated transition matrix's diagonal; duration follows equation (6). Shares may not sum to 100% because of rounding. Duration estimates for rare states have limited sample support.*

Estimated persistence is 0.857 for restrictive and rising and 0.846 for restrictive and falling. Restrictive and stable is less persistent, at 0.316. This pattern suggests that, under the chosen thresholds, directional episodes persist longer than the intermediate category. However, the 0.857 persistence estimate for accommodative and falling should not be assigned the same confidence: it comes from only seven observations concentrated in a single episode.

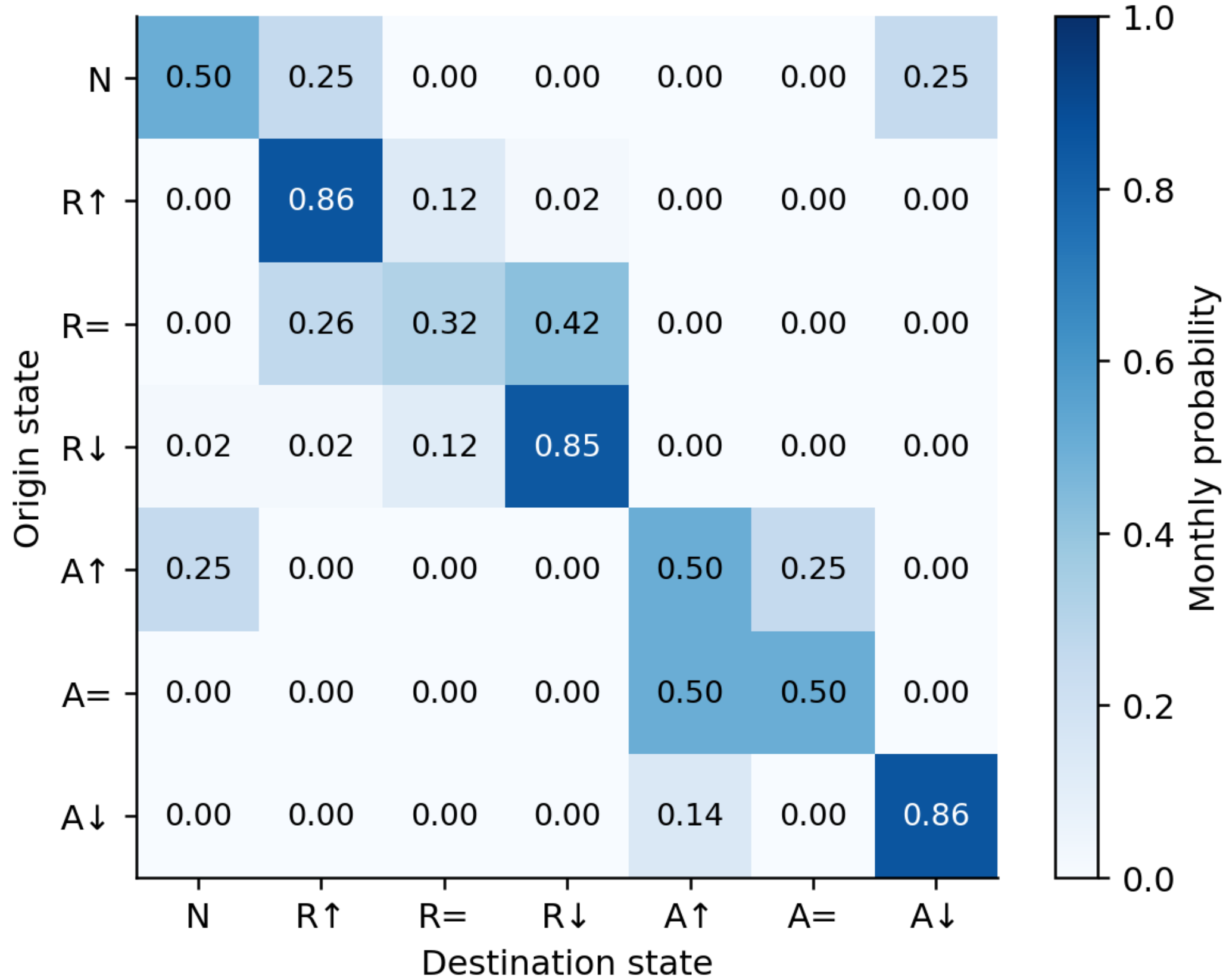

***Figure 2:*** *Transition probabilities across the seven states*

*Source: 144 observed monthly transitions. N denotes neutral transition, R restrictive, and A accommodative. Arrows indicate the direction of the gap, and the equals sign denotes stability. Zeros are unobserved transitions in this sample, not economically impossible transitions.*

The matrix is concentrated on the diagonal and on connections between restrictive states. From restrictive and falling, 44 of 52 transitions remain in the same category, and six move to restrictive and stable. The estimated stationary distribution places 87.1% of its mass on the three restrictive states. Its proximity to the observed sample share is consistent with the fitted matrix but does not independently validate either the neutral benchmark or the process's stability.

The exported global test reports a statistic of 374.24 with 36 degrees of freedom and an asymptotic p-value of approximately $7.1 \times 10^{-58}$. However, 83.7% of cells have an expected count below five. The result indicates strong association in the observed table, but its nominal significance should be treated cautiously. No cell-level inferential conclusions are reported, and the test is not interpreted as verification of the Markov assumption.

### 5.2 Latent segmentation and HMM selection

The four-state HMM has a BIC of 2,471.30, below the three-state model's 2,581.11. The composite score also favors four states. The preference is not uniform across criteria: temporal validation negative log-likelihood per observation is lower for three states, at 7.927, than for four states, at 8.142. The outputs therefore support a better penalized fit for the larger model without showing superior performance on the validation block.

**Table 3:** Comparison of HMM specifications

| States | Parameters | Log likelihood | BIC | Validation NLL | Composite score |
|---|---|---|---|---|---|
| 3 | 50 | −1,166.14 | 2,581.11 | 7.927 | 3,073.21 |
| 4 | 71 | −1,058.98 | 2,471.30 | 8.142 | 2,988.95 |

*Source: exported project comparison. The fitting sample contains 145 observations and the final validation block 29. NLL is the median across seeds; lower values are preferable. Losses are calculated on standardized variables.*

The selected segmentation contains states with 21 to 57 observations. States 0 and 2 have large average gaps but opposite average directions. State 1 groups observations with a lower average gap, while state 3 occupies an intermediate region of restraint. State identifiers have no intrinsic economic ordering.

**Table 4:** Characteristics of the four latent states

| State | Months | Mean gap | Mean 3-month change | Mean slope | Implied duration |
|---|---|---|---|---|---|
| 0 | 29 | 6.80 | −1.02 | −1.28 | 9.54 |
| 1 | 21 | −0.94 | −0.43 | 1.15 | 10.45 |
| 2 | 57 | 8.47 | 0.94 | 0.33 | 18.83 |
| 3 | 38 | 2.98 | −0.31 | 0.21 | 9.36 |

*Source: HMM summaries and transition matrix. Gaps, changes, and slopes are in percentage points; durations are in months. Means use Viterbi state assignments, whereas durations come from the estimated transition matrix.*

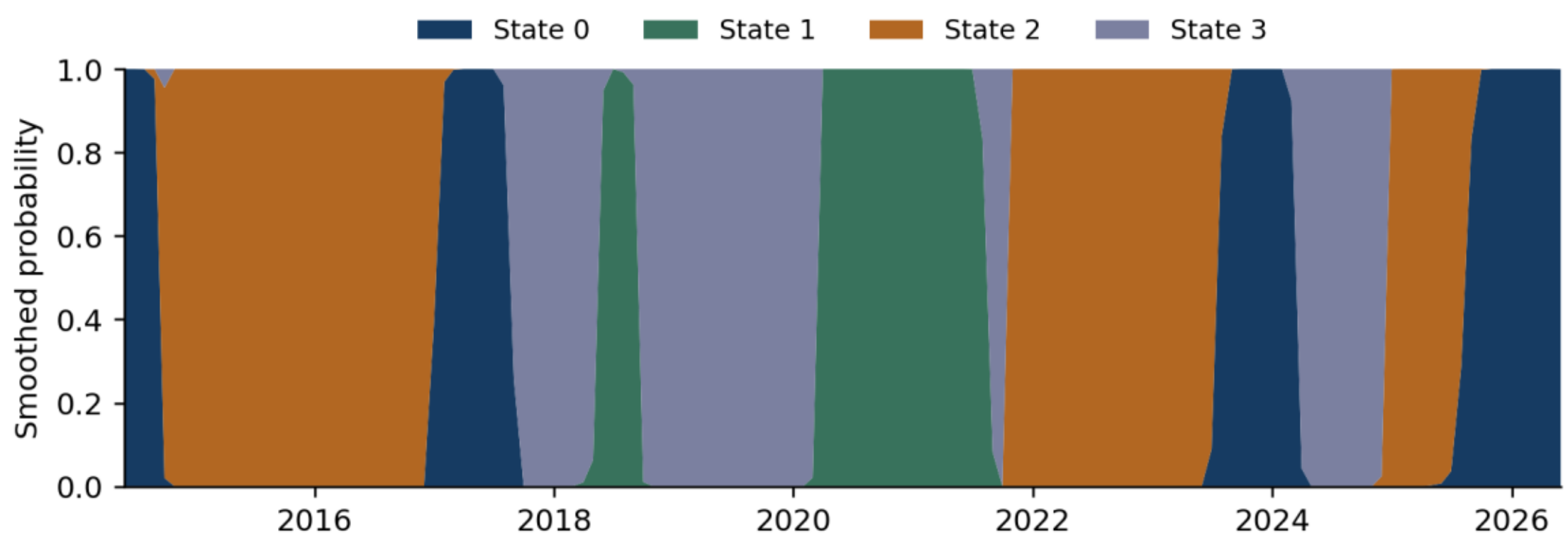

***Figure 3:*** *Smoothed probabilities of the latent states*

*Source: four-state HMM outputs. Probabilities are conditional on the full sample and do not necessarily represent the inference that would have been available at each historical date.*

Mapping each latent state to the most frequent rule label within it produces agreement in 86 of 145 months, or 59.3%. This is an in-sample descriptive correspondence, not accuracy against true labels. Because the HMM includes both variables determining the rule, agreement does not independently validate the estimated economic stance.

Interpretation must also acknowledge an algebraic restriction. Components $g_t$, $f_t$, and $s_t$ satisfy equation (2) exactly. The maximum discrepancy in the files is below $4 \times 10^{-15}$ percentage points, consistent with numerical rounding. An exact linear relation remains after standardization. The diagonal Gaussian emission ignores this dependence and assigns density to combinations that cannot occur by construction. Information criteria should therefore be read as internal criteria for this approximation. Reestimation should remove one redundant component before assessing more flexible covariance structures.

### 5.3 Selic adjustment regimes

The regression suite contains 12 combinations of specification and regime count. Four switching-variance solutions, in families B and D, are discarded for degeneracy. Their minimum variances range approximately from $10^{-33}$ to $10^{-30}$ squared percentage points, with artificially low BICs. This diagnostic demonstrates why a mechanical lowest-BIC ranking would select an unsuitable solution.

Among admissible models, specification C with three regimes has a BIC of 166.78. The same equation with two regimes yields 197.47. Specification A with three regimes, estimated in levels with inertia, yields 219.87. Selection should be understood within the suite and its admissibility conditions; it does not compare the regression block with the HMM.

**Table 5:** Regimes of the selected Selic change model

| Regime | Months | Share | Mean Selic change | Persistence | Duration |
|---|---|---|---|---|---|
| More intensive tightening | 25 | 11.4% | 0.650 | 0.216 | 1.28 |
| More intensive easing | 25 | 11.4% | −0.540 | 0.304 | 1.44 |
| Holds and gradual adjustments | 169 | 77.2% | 0.003 | 0.955 | 22.31 |

*Source: three-regime model C. Mean changes are in percentage points per month; durations are in months. Means are observed within groups assigned by their highest smoothed probability and are not the estimated regression intercepts.*

The dominant regime combines rate holds with gradual adjustments. Cuts of 0.25 percentage points in March and April 2026 are assigned to this regime, as is the unchanged rate recorded in the partial month of May. A label restricted to holds would therefore be too narrow. The other two regimes group larger average increases and decreases, with short implied durations.

Although the code exports hawkish and dovish labels, this paper uses descriptions of observed adjustments. The selected specification does not allow response coefficients to change across regimes. It therefore does not establish that the states represent structural shifts in central bank preferences or compliance with the Taylor principle. Extreme probabilities for some transitions and the smaller occupancy of the intensive-adjustment regimes also limit the precision of their interpretation.

### 5.4 Sensitivity to thresholds and the neutral benchmark

The exported grid combines four level thresholds, from 0.25 to 1.00 percentage points, and three direction thresholds, from 0.10 to 0.50 percentage points. Across all 12 combinations, the final observation remains restrictive and falling. Historical state changes range from 28 to 36, while the aggregate share of restrictive states ranges from 86.9% to 89.0%. This establishes stability of that share within the chosen grid but does not resolve uncertainty about the benchmark.

To examine the latter issue, an accounting exercise was added using the same files. In the alternative scenario, the supplied series is interpreted as a real neutral rate $r_t^*$ and converted into a nominal benchmark using 12-month expected inflation:

$$n_t^{\text{alt}} = 100\left[\left(1+\frac{r_t^*}{100}\right)\left(1+\frac{\pi_t^e}{100}\right)-1\right], \qquad g_t^{\text{alt}} = i_t - n_t^{\text{alt}}. \qquad (12)$$

The same rule is reapplied to the 145 months with unchanged thresholds and direction horizon. This is not asserted to be the correct interpretation of the series; the exercise measures dependence on an unconfirmed source convention. The HMM and regressions were not reestimated under this scenario.

**Table 6:** Sensitivity to the interpretation of the neutral benchmark

| Measure | Series treated as nominal | Series treated as real |
|---|---|---|
| Classified months | 145 | 145 |
| Restrictive share | 88.3% | 51.0% |
| Neutral share | 2.8% | 9.7% |
| Accommodative share | 9.0% | 39.3% |
| State changes | 35 | 40 |
| Final-month gap | 5.55 pp | 1.14 pp |
| Final-month state | Restrictive and falling | Restrictive and falling |

*Source: calculations using the supplied series, expected inflation, and the project's rule. The second scenario is a sensitivity exercise added during preparation of this paper. The final month is partial.*

The alternative interpretation changes the label of 74 observations, or 51.0% of the sample. The restrictive share falls by 37.2 percentage points. This effect is much larger than the variation associated with the threshold grid. Establishing the economic definition of the neutral series is therefore more consequential for the historical composition of states than small changes in the rule's bands.

In the final month, direction remains negative in both scenarios. Under the original convention, the gap is 5.55 percentage points and its three-month change is −1.15 percentage points; under the alternative, the corresponding values are 1.14 and −1.32 percentage points. April, the last complete month in the financial dataset, is already restrictive and falling under the original convention, with a gap of 5.80 percentage points and a change of −1.21 percentage points. The terminal rule classification does not depend solely on the partial month, although the magnitude of restraint depends on the neutral rate convention.

## 6 Discussion and implications for monetary monitoring

The framework's interpretive benefit lies in separating its objects of analysis. The rule describes the relative position of interest rates and the movement of their gap. The HMM identifies statistical groupings of the selected variables. The preferred regression distinguishes conditional patterns in monthly Selic changes. Joint interpretation is useful when these differences are preserved; seeking a single consensus regime across models that address different questions would create a false equivalence.

This separation allows a situation to be described as still restrictive, with a falling interest rate gap and gradual nominal adjustments. The three descriptions are consistent with one another. Likewise, an inverted curve does not imply that policy has already become accommodative: the curve measures relative prices across maturities, whereas the classification depends on the neutral benchmark.

The application's limitations follow an economic and statistical order. First, the neutral series must be documented, including whether it is real or nominal, its horizon, conversion equation, components, and availability dates. The sensitivity in Table 6 prevents the predominance of restrictive states from being treated as an unconditional fact about the period. For real-time interpretation, it also matters whether the series was estimated using information that lay in the future relative to each month.

Second, the HMM variable set needs to be reduced to respect the identity linking the gaps and the slope. Retaining redundant information under diagonal emissions affects the relative weight of interest rates in the likelihood. Comparing nonredundant variable sets, diagonal covariances, and regularized covariance structures would help distinguish economic information from the geometry imposed by the representation.

Third, the empirical chain requires tests of stability, alternative orders, and dependence of duration on elapsed time in a state. Overlapping changes and aggregation into seven categories may generate persistence even when the underlying process is not truly a homogeneous Markov chain. For rare states, uncertainty intervals based on methods that respect temporal dependence and sensitivity exercises would be more informative than isolated point probabilities.

Fourth, temporal evaluation should reconstruct variables within each window. This includes the activity gap proxy, neutral rate estimates, hyperparameter selection, and actual release dates. An expanding-window evaluation could compare predicted probabilities with subsequently observed states using log loss and the Brier score, alongside simple benchmarks such as persistence and historical frequencies. These exercises are proposed extensions and were not performed in this paper.

Finally, the Selic change equation should be assessed in light of the discrete nature and calendar of Copom decisions. The contemporaneous slope contains the Selic rate itself and may reflect simultaneity, precluding a causal interpretation of its coefficient. A model at meeting frequency, with information observed before the decision and a distribution compatible with increases, holds, and cuts, would be more directly aligned with monetary decisions. Excluding degenerate variances addresses an important numerical pathology but does not resolve these specification issues.

## 7 Conclusion

The project provides an interpretable framework for organizing monthly assessments of Brazilian monetary policy. Its central feature is the separation of the level of the interest rate gap from its direction, followed by comparison with latent states and Selic adjustment regimes. The observable rule makes classifications traceable; probabilistic models add a description of persistence and historical heterogeneity.

The examined files support three conditional findings. Under the benchmark treated as nominal, restrictive states predominate, and directional states are more persistent than the restrictive stable state. The four-state HMM improves BIC and the composite score but not the available temporal validation loss. Regressions distinguish more intensive increases and decreases from a broader regime of holds and gradual adjustments after numerically degenerate solutions are excluded.

Sensitivity to the interpretation of the neutral rate substantially changes the historical distribution, while the terminal classification remains restrictive and falling in both scenarios. The most important substantive implication is therefore the need to tie regime interpretation to a verifiable definition of neutrality. Documenting that input, removing HMM redundancy, and reconstructing a complete temporal evaluation are the next steps toward turning the historical description into a predictive tool with measurable performance.

## Data and code availability

This paper was revised on July 31, 2026. Preserved Markov, HMM, and regression outputs were checked against the source code. The classification, transition counts, HMM-to-rule agreement, and sensitivity to the nature of the neutral rate were recalculated. Latent models were not reestimated.

## Appendix A Regression specifications

**Table A1:** Structure of the switching regression suite

| Model | Dependent variable | Regressors | Switching parameters |
|---|---|---|---|
| A | Selic | Inflation, activity gap, FX, lagged Selic | Intercept |
| B | Selic | Inflation, activity gap, FX, lagged Selic | Intercept and variance |
| C | Selic change | Inflation, activity gap, FX, slope | Intercept |
| D | Selic change | Inflation, activity gap, FX, slope | Intercept and variance |
| E | Selic | Inflation, activity gap, FX | Intercept and variance |
| F | Selic | Inflation, activity gap, lagged Selic | Intercept and coefficients |

*Note: inflation denotes the deviation of the relevant-horizon projection from the interpolated target; FX denotes the three-month log exchange rate change. All families are estimated with two and three regimes.*

**Table A2:** Selection criteria and admissibility of switching regressions

| Model | Regimes | BIC | Convergence | Assessment |
|---|---|---|---|---|
| A | 2 | 249.70 | No | Exploratory only |
| A | 3 | 219.87 | Yes | Admissible |
| B | 2 | −7,762.43 | No | Degenerate variance |
| B | 3 | −7,924.18 | No | Degenerate variance |
| C | 2 | 197.47 | Yes | Admissible |
| C | 3 | 166.78 | Yes | Selected |
| D | 2 | −8,555.72 | Yes | Degenerate variance |
| D | 3 | −8,486.65 | Yes | Degenerate variance |
| E | 2 | 887.66 | Yes | Admissible |
| E | 3 | 796.87 | Yes | Admissible |
| F | 2 | 224.65 | Yes | Admissible |
| F | 3 | 236.16 | Yes | Admissible |

*Source: exported comparison table. BIC values for degenerate solutions document the failure and are excluded from selection. Numerical convergence alone does not establish admissibility.*

## Appendix B Traceability of results

Classifications and gaps come from monthly_monetary_quadrant_dataset.csv. Frequencies, transitions, and chain diagnostics are in processed/markov. HMM comparisons, states, probabilities, and parameters are in processed/hmm. Regressions, diagnostics, and regime summaries are in processed/taylor_ms. Main-text tables use these outputs, except for the alternative scenario in Table 6, recalculated using equation (12).

Arithmetic checks reproduced all 145 rule labels, 144 transitions, the identities in equations (1) and (2), the 86 months of mapped agreement, and the neutral rate sensitivity results. They establish consistency with the supplied files, not reestimation of the latent models or external validation of the neutral benchmark.